\documentclass{aa}  

\usepackage{graphicx}
\usepackage{txfonts}
\usepackage{xspace}
\usepackage{rotating}
\usepackage{pdflscape}
\usepackage{longtable}
\usepackage{booktabs}
\usepackage{color}
\usepackage{url}
\usepackage{hyperref}

\usepackage{xcolor}
\hypersetup{
  colorlinks   = true, 
  urlcolor     = blue, 
  linkcolor    = blue, 
  citecolor   = blue 
}
\usepackage{siunitx}
\usepackage{rotating}
\newcommand{\Nu}{{\it NuSTAR\xspace}}
\newcommand{\ch}{{\it Chandra\xspace}}

\usepackage[flushleft]{threeparttable}
\usepackage{graphicx,epstopdf}
\DeclareGraphicsExtensions{.ps}
\def\Msun{M_\odot}

\def\flux{erg s$^{-1}$ cm$^{-2}$}

\graphicspath{{./}{figures/}}

\begin{document}

\title{A search for high-redshift direct-collapse black hole candidates in the PEARLS north ecliptic pole field}

\author{Armin Nabizadeh \inst{1} 
        \and Erik Zackrisson \inst{1,22}
        \and Fabio Pacucci \inst{2}
   	\and W. Peter Maksym \inst{2,24}
   	\and Weihui Li \inst{1}
        \and Francesca Civano\inst{2}
        \and Seth H. Cohen\inst{3} 
        \and Jordan C. J. D'Silva\inst{8,9} 
        \and Anton M. Koekemoer\inst{4} 
        \and Jake Summers\inst{3} 
        \and Rogier A. Windhorst\inst{3} 
        \and Nathan Adams \inst{7}
        \and Christopher J. Conselice\inst{7} 
        \and Dan Coe\inst{4,5,6} 
        \and Simon P. Driver\inst{8} 
        \and Brenda Frye\inst{10} 
        \and Norman A. Grogin\inst{4} 
        \and Rolf A. Jansen\inst{3} 
        \and Madeline A. Marshall\inst{11,9} 
        \and Mario Nonino\inst{12} 
        \and Nor Pirzkal\inst{4} 
        \and Aaron Robotham\inst{8} 
        \and Michael J. Rutkowski\inst{13} 
        \and Russell E. Ryan, Jr.\inst{4} 
        \and Scott Tompkins\inst{3} 
        \and Christopher N. A. Willmer\inst{10} 
        \and Haojing Yan\inst{14} 
        \and Jose M. Diego\inst{15,16}
        \and Cheng Cheng\inst{17}
        \and Steven L. Finkelstein\inst{18}
        \and S. P. Willner\inst{2}
        \and Lifan Wang\inst{23}
        \and Adi Zitrin\inst{19}
        \and Brent M. Smith\inst{3}
        \and Rachana Bhatawdekar\inst{20}
        \and Hansung B. Gim\inst{21}
          }

\institute{Observational Astrophysics, Department of Physics and Astronomy, Uppsala University, Box 516, SE-751 20 Uppsala, Sweden \\ \email{armin.nabizade@gmail.com} 
\and
Center for Astrophysics \textbar\ Harvard  \& Smithsonian, 60 Garden St., Cambridge, MA 02138, USA 
\and
School of Earth and Space Exploration, Arizona State University, Tempe, AZ 85287-1404, USA 
\and
Space Telescope Science Institute, 3700 San Martin Drive, Baltimore, MD 21218, USA 
\and
Association of Universities for Research in Astronomy (AURA) for the European Space Agency (ESA), STScI, Baltimore, MD 21218, USA 
\and
Department of Physics, 366 Physics North MC 7300, University of California, Berkeley, CA 94720, USA 
\and
Jodrell Bank Centre for Astrophysics, Alan Turing Building, University of Manchester, Oxford Road, Manchester M13 9PL, UK 
\and
International Centre for Radio Astronomy Research (ICRAR) and the International Space Centre (ISC), The University of Western Australia, M468, 35 Stirling Highway, Crawley, WA 6009, Australia 
\and
ARC Centre of Excellence for All Sky Astrophysics in 3 Dimensions (ASTRO 3D), Australia 
\and
Department of Astronomy/Steward Observatory, University of Arizona, 933 N Cherry Ave, Tucson, AZ, 85721-0009, USA 
\and
National Research Council of Canada, Herzberg Astronomy \& Astrophysics Research Centre, 5071 West Saanich Road, Victoria, BC V9E 2E7, Canada 
\and
INAF-Osservatorio Astronomico di Trieste, Via Bazzoni 2, 34124 Trieste, Italy 
\and
Minnesota State University-Mankato,  Telescope Science Institute, TN141, Mankato MN 56001, USA 
\and
Department of Physics and Astronomy, University of Missouri, Columbia, MO 65211, USA 
\and
Instituto de Física de Cantabria, Edificio Juan Jordá, Avenida de los Castros s/n, E-39005 Santander, Cantabria, Spain 
\and
Instituto de Física de Cantabria (CSIC-UC). Avenida. Los Castros, s/n. E-39005 Santander, Spain 
\and
Chinese Academy of Sciences, National Astronomical Observatories, CAS, Beijing 100101, People's Republic of China 
\and
Department of Astronomy, The University of Texas at Austin, 2515 Speedway, Stop C1400, Austin, Texas 78712, USA 
\and
Physics Department, Ben-Gurion University of the Negev, P.O. Box 653, Beer-Sheva 8410501, Israel 
\and
European Space Agency (ESA), European Space Astronomy Centre (ESAC), Camino Bajo del Castillo s/n, 28692 Villanueva de la Cañada, Madrid, Spain 
\and
Department of Physics, Montana State University, P. O. Box 173840, Bozeman, MT 59717, USA 
\and Swedish Collegium for Advanced Study,Linneanum
Thunbergsv\"a{}gen 2, SE-752 38 Uppsala, Sweden 
\and Mitchell Institute for Fundamental Physics and Astronomy, Texas A\&M University, College Station, TX 77843, USA 
\and NASA Marshall Space Flight Center, Huntsville, AL 35812, USA  
}
          
\titlerunning{A Search for DCBHs in the NEP}
\authorrunning{Nabizadeh et al.}

   \date{2023}

\abstract{Direct-collapse black holes (DCBHs) of mass $\sim 10^4$--$10^5 \Msun$ that form in HI-cooling halos in the early Universe are promising progenitors of the $\gtrsim 10^9 \Msun$ supermassive black holes that fuel observed $z \gtrsim 7$ quasars. Efficient accretion of the surrounding gas onto such DCBH seeds may render them sufficiently bright for detection with the JWST up to $z\approx 20$. Additionally, the very steep and red spectral slope predicted across the $\approx 1$--5 $\mu$m wavelength range of the JWST/NIRSpec instrument during their initial growth phase should make them photometrically identifiable up to very high redshifts. In this work, we present a search for such DCBH candidates across the 34 arcmin$^{2}$ in the first two spokes of the JWST cycle-1 PEARLS survey of the north ecliptic pole time-domain field covering eight NIRCam filters down to a maximum depth of $\sim$ 29\,AB\,mag. We identify two objects with spectral energy distributions consistent with theoretical DCBH models. However, we also note that even with data in eight NIRCam filters, objects of this type remain degenerate with dusty galaxies and obscured active galactic nuclei over a wide range of redshifts. Follow-up spectroscopy would be required to pin down the nature of these objects. Based on our sample of DCBH candidates and assumptions on the typical duration of the DCBH steep-slope state, we set a conservative upper limit of  $\lesssim 5\times 10^{-4}$\,comoving Mpc$^{-3}$ (cMpc$^{-3}$) on the comoving density of host halos capable of hosting DCBHs with spectral energy distributions similar to the theoretical models at $z\approx 6$--14. 
}


\maketitle
\section{Introduction}
\label{sec:intro}
In the history of the Universe, the evolution of galaxies and supermassive black holes (SMBHs) are tightly connected. More than 200 quasars powered by SMBHs of mass $\gtrsim 10^9 \Msun$ have been discovered at redshift $z\gtrsim 6$ \citep[e.g.][]{Fan_2001, Fan_2003, Mortlock_2011, Wu_2015, Banados_2018, Yang_2020, Wang21, Fan_2023_review}, with some of them shining when the age of the Universe was less than 800 Myr. Explaining how black holes reach such masses this early in the history of the Universe is challenging and requires some combination of highly efficient gas accretion and black hole mergers \citep{Pacucci_2020} starting from lower-mass ($\sim 10^2$--$10^6 \Msun$) black hole seeds \citep[see, e.g.,][for recent reviews]{Woods19, Inayoshi_review_2019, Fan_2023_review}. Thus, the discovery of high-redshift SMBHs is instrumental in constraining the properties of the seed population of black holes \citep{Pacucci_2022_search}.

Several formation mechanisms for such seeds have been proposed in the literature. These include (i) primordial formation during inflation or from cosmic string loops \citep[e.g.,][]{Hasinger_2020}; (ii) formation as the end product of massive and metal-free stars, possibly undergoing super-Eddington accretion episodes \citep[e.g.][]{Begelman_1978, Wyithe_Loeb_2012, Begelman_Volonteri_2017}; (iii) formation through runaway stellar mergers in young star clusters \citep[e.g.][]{PZ_2002, Davies_2011, Katz_2015, Boekholt_2018}; and (iv) formation via direct-collapse black holes \citep[DCBHs;][]{Loeb_Rasio_1994, Bromm_Loeb_2003, Lodato_Natarajan_2006, Pacucci_2017, Inayoshi_review_2019, Volonteri21}. 

In the DCBH scenario, gas within an atomic-cooling halo (i.e., $\gtrsim 10^7\,\Msun$) collapses to form a $\sim 10^4$--$10^6\,\Msun$ \citep{Ferrara_2014} black hole at $z\sim 20-$30 \citep[e.g.][]{Loeb_Rasio_1994, Bromm_Loeb_2003, Lodato_Natarajan_2006}, possibly with a supermassive star or a quasi-star formed as an intermediate step \citep[see][for a review]{Woods19}. Accretion onto the newly formed seed black hole from the surrounding gas in duty cycles with active phases lasting up to $\approx 100$\,Myr \citep[e.g.][]{Pacucci15} could render such a DCBH sufficiently luminous to allow detection at $z>7$ either in the near-to-mid infrared (IR) or in the X-ray regime. The high-energy emission is more sensitive to details of the accretion process, such as the gas metallicity and column density of the host galaxy \citep{Pacucci15, Pacucci16}. Current X-ray observatories, such as Chandra, have deep-field sensitivity limits of $\sim 10^{-17} \, \rm erg \, s^{-1} \, cm^{-2}$ that render most of the seed population undetectable. Proposed X-ray probe-class missions, such as AXIS, could reach sensitivities of $\sim 10^{-18} \, \rm erg \, s^{-1} \, cm^{-2}$ in their deep fields, allowing the uncovering of at least part of the population of seeds, especially in the heavy regime.
In the deepest James Webb Space Telescope (JWST) exposures, DCBHs of initial mass $\sim 10^5 \Msun$ may remain detectable up to $z\approx 20$ \citep{Natarajan17,Whalen20} if supplied with large accretion rates from their host.

The likely observable signatures of DCBHs through JWST observations vary significantly, largely due to various factors as discussed in the literature \citep{Pacucci15, Pacucci16, Natarajan17, Valiante17, Visbal_2018, Valiante18, Whalen20, Inayoshi_2022, Nakajima_2022}. These factors include (i) the evolutionary state of the objects, including factors such as the initial seed mass, gas density, and metallicity of the host, as well as gas availability; (ii) the details of the accretion process onto the seed (e.g., the geometry of the disk and its radiative efficiency); (iii) the triggering of star formation in the host, possibly enhanced by soft X-ray irradiation from the DCBHs; and (iv) the merger history of DCBHs after formation. 

Some models suggest that the light from the DCBH is likely to be blended with that of surrounding stars and associated nebular emission \citep{Natarajan17}. This occurs either because the DCBH forms in the direct vicinity of star-forming halos (which provides the radiation required to deplete molecules in the DCBH host and prevent cooling) and merges with these on timescales as short as $\sim 1$\,Myr \citep[e.g.][]{Pacucci_2017, Natarajan17}, or because star formation occurs within the DCBH host halo itself \citep[e.g.][]{Aykutalp_2014, Valiante18, Barrow18, Aykutalp20}. In other scenarios, the DCBH can remain isolated for prolonged periods while efficiently growing in mass due to cold accretion \citep{Whalen20, Latif22}. 

Consequently, the anticipated photometric spectral energy distributions (SEDs) of high-redshift DCBHs, spanning the wavelength range probed by JWST exhibit a spectrum ranging from blue \citep[e.g.][]{Valiante18,Barrow18} to red \citep[e.g.][]{Pacucci16,Whalen20,Inayoshi22}. The predicted formation rates of DCBHs as a function of redshift are also highly variable, mainly due to uncertainties in the level of background radiation necessary to prevent star formation and the effect of supernova feedback \citep{Habouzit_2016}. Theoretical predictions on the comoving number densities of halos capable of hosting DCBHs span over five orders of magnitude at any given redshift \citep{Habouzit_2016, Valiante17}, rendering estimates of DCBH detectability with JWST highly uncertain \citep[e.g.,][]{Pacucci_2019_BAAS}.

This work presents a photometric search for $z\gtrsim 6$ DCBH candidates in the Prime Extragalactic Areas for Reionization and Lensing Science (PEARLS) north ecliptic pole (NEP) field for which auxiliary X-ray data are available. In particular, we focus on the predictions of \citet{Pacucci16} in which DCBHs, at birth, exhibit very steep (i.e., "red'') SEDs within the wavelength range of $\approx 1$--5 $\mu $m probed by JWST/NIRCam. These SED models allow us to efficiently sift out promising candidates with current JWST observations because of the large change in flux between the near-to-mid IR filters. 

A previous search by \citet{Pacucci16} for such red DCBH candidates at $z\lesssim 10$ using data from the Hubble Space Telescope (HST), Spitzer, and \ch\ revealed that objects at $H<27$\,AB\,mag exhibiting the relevant $\approx 1$--5$\mu $m signatures of $\sim 10^4$--$10^6\,\Msun$ DCBHs do exist but are very rare. 
However, this candidate selection was based on three broadband filters only in the range of 1--5 $\mu$m. Given the superior depth and more diverse set of photometric filters provided by JWST, it remains unclear how many of such candidates will remain consistent with a DCBH interpretation after further scrutiny.

The current work is organized as follows. In Section \ref{sec:PEARLS}, we present the PEARLS data used, while Section~\ref{sec:SED} provides a description of the selection criteria upon which our search is based. In Section~\ref{sec:model-candidates}, we present our main results: our DCBH candidates and the inferred upper limits on the comoving number density of DCBH host halos. Our findings are discussed and summarized in Section~\ref{sec:discussion}.

\section{JWST/NIRCam data on the NEP field}
\label{sec:PEARLS}
The IR observations analyzed in this study were provided as a part of the  PEARLS \citep[PI: R. Windhorst;][]{Windhorst2023} Guaranteed Time Observations (GTO) program. PEARLS is a time-domain survey of the NEP field that is being carried out in four "spokes" \citep{Jansen18} covering a total survey area of 68\,arcmin$^{2}$ \citep[i.e., 17\,arcmin$^{2}$ for each spoke as a $2 \times 1$ mosaic layout of NIRCam modules A and B;][]{Rieke2023-NIRCam}. The NEP time-domain field, centered at RA 17:22:47.896 and Dec +65:49:21.54, is within the JWST continuous viewing zone and has been imaged by the NIRCam in eight near-IR bands using four short-wavelength (SW) filters (F090W, F115W, F150W, and F200W) and four long-wavelength (LW) filters (F277W, F356W, F410M, and 444W) with $5\sigma$ detection limits at 28--29\,AB\,mag. The auxiliary data for this field contains observations spanning a broad energy range from radio to X-ray \citep[see~][for additional details]{Windhorst2023}. In this work, we used spoke 1 and spoke 2 data catalogs covering a total area of 34 arcmin$^2$ and including $\approx 24120$ objects. 

The catalogs were created following the methodology outlined in \citet{Windhorst2023}, with the only significant difference being the retention of point sources, as it is possible that DCBHs remain unresolved. In short, Source Extractor \citep{1996A&AS..117..393B} was utilized for source detection on 30 milli-arcsecond pixel-scale mosaics astrometrically aligned to Gaia DR3 and with weight maps employed to aid in the detection and to account for flux uncertainties. The dual-image mode was employed, with the F444W image used for source detection and aperture definition. The minimum threshold for analysis and detection was set at 1.5$\sigma$ in nine connected pixels, while the deblending contrast parameter was set to 0.06 to strike a balance between completeness and reliability. All measured fluxes used Source Extractor's "automatic" apertures (i.e., \emph{MAG\_AUTO}).

Additionally, another important dataset in our search for $z>6$ DCBH candidates was the \Nu\ 3--24\,keV and \ch\ 0.5--7\,keV X-ray observations. We used the dataset to select X-ray bright candidates (see Sec.~\ref{sec:X-ray}).

\section{Photometric signatures of direct-collapse black holes at $z\gtrsim 6$}
\label{sec:SED}
High-redshift accreting DCBHs that are surrounded by a sufficiently dense reservoir of gas are Compton-thick (i.e., they have a column density $n_H > 1.5 \times 10^{24} \, \rm cm^{-2}$) and can display a very red SED for prolonged periods of time at rest-frame ultraviolet and optical wavelengths \citep{Pacucci16}. This serves as a photometric signature across the 1--5 $\mu$m range that is probed by JWST/NIRCam at $z\gtrsim 6$. In this phase of evolution, the DCBH may also be detectable in the X-ray regime, although the X-ray flux is highly model dependent and can lie below the detection threshold of all current X-ray telescopes \citep{Pacucci15}, except during the brightest phases of accretion.

\begin{figure}
\begin{center} 
\includegraphics[width=\columnwidth]{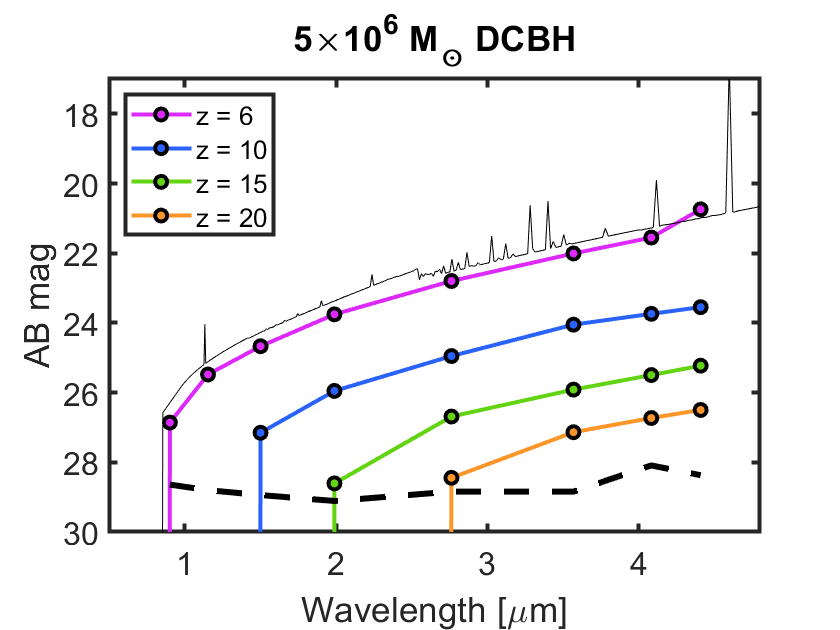}
\end{center}
\caption{DCBH detection limits in the NEP. The thick dashed line represents the combined JWST/NIRCam $5\sigma$ flux detection limits expressed in AB magnitudes across the NEP spokes.  The colored lines show the photometric fluxes of the \citet{Pacucci16} $5\times 10^6 \Msun$ DCBH model at $z=6$, 10, 15, and 20 in the set of NIRCam filters used by PEARLS across the NEP field. The thin solid black line shows the $z=6$ version of the DCBH model spectrum on which the photometric predictions are based, but offset by 0.5 magnitudes for clarity. As can be seen, this \citet{Pacucci16} $5\times 10^6 \Msun$ DCBH model remains detectable up to $z=20$ in the reddest NIRCam filters used. The photometric SEDs are characterized by a significantly red slope across the NIRCam bands and are relatively featureless except for the drop at the Ly$\alpha$ limit and the bump seen in the F444W filter at $z=6$ SED, which is due to strong H$\alpha$ emission. This is not seen in the higher-redshift SEDs, since it redshifts out of the NIRcam range at $z\gtrsim 6.9$.}
\label{fig:SED} 
\end{figure}

\begin{figure}
\begin{center} 
\includegraphics[width=\columnwidth]{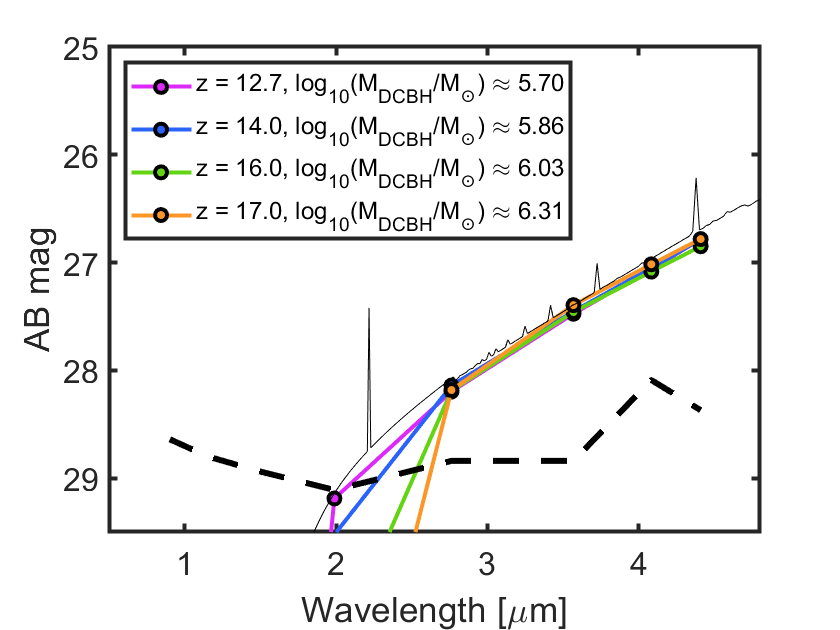}
\end{center}
\caption{Mass-redshift degeneracy for DCHB SEDs without a clear-cut Ly$\alpha$ break. If the Ly$\alpha$ break is close to or below the detection threshold (dashed black line), acceptable SED fits to NIRCam data at longer wavelengths can be achieved for DCBH models over a wide range of redshifts. Here, the purple line corresponds to the photometric SED of a $z=12.7$ DCBH with mass $\log_{10}(M_\mathrm{DCBH}/M_\odot)=5.70$, and the thin solid line is the corresponding spectrum (offset by 0.1 magnitudes for clarity). In this case, the Ly$\alpha$ break occurs at $\approx 1.7\ \mu$m between the F150W and F200W filters. While the F200W flux is largely unaffected by the Ly$\alpha$ break, the red slope of the DCBH SED still places the F200W flux  below the detection limit of the NEP field. Since a sharp drop in flux at the Ly$\alpha$ limit becomes unobservable in this case, the redshift becomes poorly constrained as higher-mass DCBH models at $z=14$ (blue line), $z=16$ (green line), and $z=17$ (orange line) produce very similar fluxes in the longer-wavelength NIRCam filters (F277W, F356W, F410M, and F444W).}
\label{fig:mass_z_degeneracy} 
\end{figure}

In Figure~\ref{fig:SED}, we plot the SED of the \citet{Pacucci16} $5\times 10^6 \Msun$ DCBH model at $z=6$--20 against the PEARLS JWST/NIRCam detection limits. As can be seen, this model remains detectable in at least four JWST/NIRCam filters until $z=20$ and displays a very red spectral slope.

While this significantly red SED ensures that DCBHs stand out from the majority of $z\gtrsim 6$ objects, it also presents a number of challenges. The intrinsically red slope of the SED can, in fact, make it difficult to assess the wavelength of the Lyman-$\alpha$ (Ly$\alpha$) break since the flux in the filter covering wavelengths directly longward of the "dropout'' filter may also fall below the detection threshold in the case of faint DCBHs. As a reminder, the Ly$\alpha$ break renders the flux at wavelengths $\lesssim 0.1216\,(1+z)\ \mu$m undetectably low due to absorption in the neutral intergalactic medium for light sources at $z\gtrsim 6$ and is often used to determine the redshift. The situation is illustrated in Figure~\ref{fig:mass_z_degeneracy}, where we show that DCBHs over a relatively large redshift range ($\Delta(z)\approx 4$) can produce photometric SEDs that are observationally indistinguishable given the NEP detection limit. 

\begin{figure}
\begin{center} 
\includegraphics[width=\columnwidth]{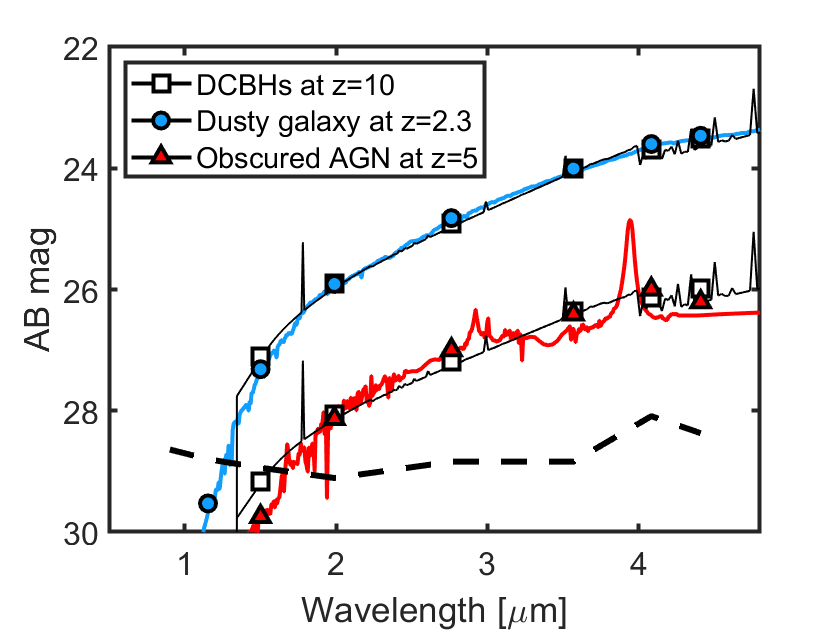}
\end{center}
\caption{Photometric degeneracy between high-$z$ DCBHs, obscured AGN, and very dusty galaxies. The thin black lines represent the \citet{Pacucci16} model spectra of DCBHs with mass $5\times 10^6\ M_\odot$ (upper line) and $5\times 10^5\ M_\odot$ (lower line) at $z=10$. The white squares indicate the corresponding integrated fluxes in the NIRCam filters used in the PEARLS/NEP survey. While the detailed spectra of the DCBHs display features that would set them apart from both dusty galaxies and obscured AGN, their photometric properties can be relatively well reproduced by both of these types of objects. The blue line represents the model spectrum of a passively evolving, very dusty ($A_\mathrm{V}\approx 2.5$\,mag, $Z=0.004$, age $\approx 2$\,Gyr) galaxy at $z=2.3$ based on the \citet{Zackrisson11a} set, with filter fluxes (blue squares) that can roughly reproduce the photometric data points of the $5\times 10^6\ M_\odot$ DCBHs that lie above the PEARLS/NEP detection threshold (dashed black line). The red line and red triangles represent the \citet{Polletta07}  template spectrum and its corresponding NIRCam filter fluxes for an obscured AGN (QSO2 template), redshifted to $z=5$ and scaled to match the $5\times 10^5\ M_\odot$ DCBH model as closely as possible. Also in this case, there is a substantial similarity in the photometric SED between the $z=5$ AGN and the $z=10$ DCBH.}
\label{fig:SED comparison} 
\end{figure}

A second challenge comes from the fact that the relatively featureless photometric SEDs of the \citet{Pacucci16} DCBHs may be confused with other rare types of objects. Low-temperature stars or substellar objects in the Milky Way could appear as very red point sources, and in some cases, they can potentially be confused with compact objects at high redshifts. However, the detailed JWST/NIRSpec SEDs of our DCBH models at $z=5-$20 do not match the SEDs of any of the faint star or brown dwarf models in the sets of \citet{Baraffe15} and \citet{Phillips20}, and such objects are therefore unlikely to be confused with DCBHs in the current search. Dusty galaxies and obscured active galactic nuclei (AGN) are more likely to be interlopers, and in Fig.~\ref{fig:SED comparison} we show that even when imaging in the eight NIRCam filters is available, the photometric signatures of DCBHs may be reproduced by both of these types of objects. When only parts of the NIRCam wavelength range are probed (i.e., when colors featuring just a few of the NIRCam wide-band filters are used as diagnostics), DCBHs may also display colors similar to some of the very reddest objects detected so far by JWST, including AGN candidates and so-called HST-dark galaxies \citep{Rodighiero_2023, Furtak_2023, Larson_2023, Barrufet_2023, Kokorev_2023, Barro2023arXiv, Matthee23, Labbe23, Smail2023}. For example, \citet{Kocevski2023} has reported on the discovery of a broad-line AGN at $z\approx 5.6$, which at $\approx 2-4.4\ \mu$m displays a steep and red SED very similar to our DCBH models, whereas the $\approx 1.15-$1.5 part is relatively flat in $f_\nu$ units \citep[many similar SEDs have also been seen in][]{Barro2023arXiv, Labbe23}. In its entirety, this type of SED would not allow for a good fit to our DCBH models, but if the source had been intrinsically fainter so that the short-wavelength part had fallen below the detection threshold, our search procedure would probably have identified this SED as a likely DCBH. 

This means that all photometrically selected DCBH candidates should be considered tentative until confirmed through spectroscopy, as discussed in Section~\ref{sec:discussion}. In Appendix~\ref{Appendix-A}, we provide some further exploration of the properties of galaxies capable of mimicking the SEDs of DCBHs.

\section{Direct-collapse black hole candidates} 
\label{sec:model-candidates}
To produce a grid of SED models for DCBHs against which the objects in the PEARLS/NEP catalogs are tested, we started from the \citet{Pacucci16} models of a DCBH with a seed mass around $10^5 \ M_\odot$, and we extracted SEDs at times where the black hole mass has grown to $M_\mathrm{DCBH} = 5\times 10^5$, $1\times 10^6$, $5\times 10^6$, and $7\times 10^6 \ M_\odot$. These four original DCBH model spectra were then interpolated to form 50 DCBH spectra uniformly spaced in $\log_{10} M_\mathrm{DCBH}$ throughout the $5\times 10^5$--$7\times 10^6\ M_\odot$ range. On this basis, we derived photometric fluxes in the relevant NIRCam filters at $z=5$--15 with a step size of $\Delta(z)=0.1$. We assumed a cosmology characterized by $\Omega_\mathrm{M}=0.32$, $\Omega_\Lambda = 0.68$, and $H_0=67$\,km\,s$^{-1}$\,Mpc$^{-1}$, as well as complete absorption by the neutral intergalactic medium at rest wavelengths shorter than Ly$\alpha$ for sources at $z>5.8$.

\subsection{Direct-collapse black hole candidate selection} 
\label{sec:candidate}
\begin{figure*}
\begin{center} 
\includegraphics[width=1.7\columnwidth]{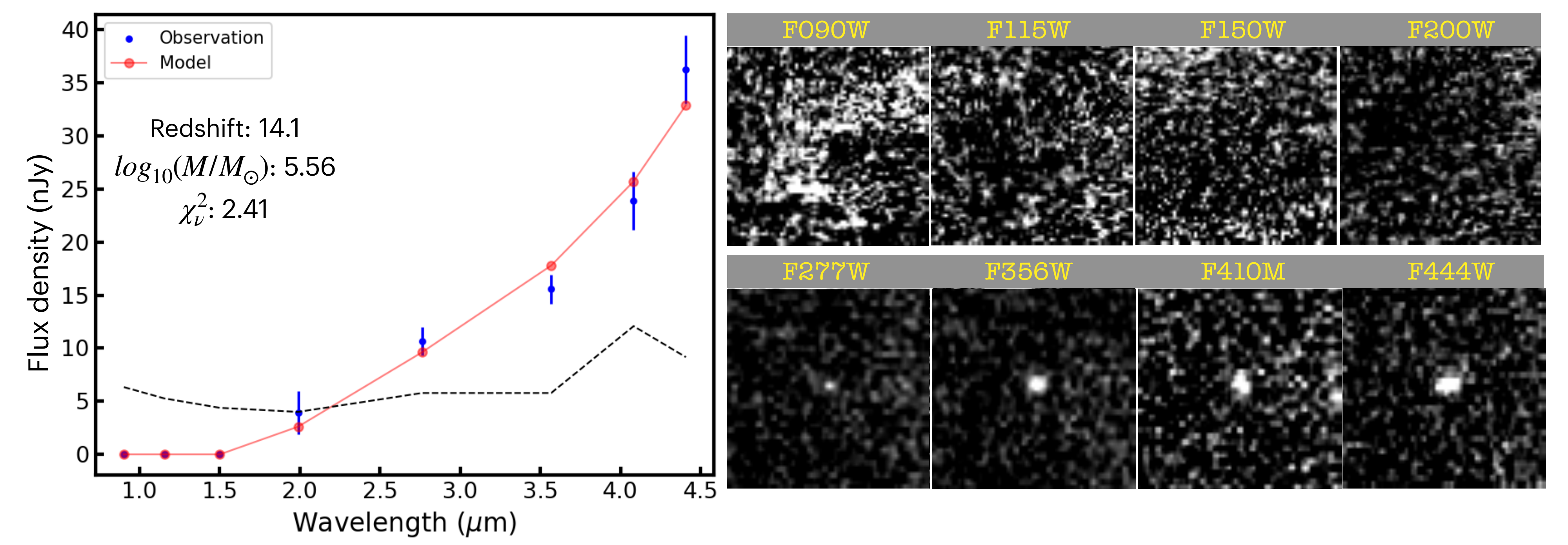}
\includegraphics[width=1.7\columnwidth]{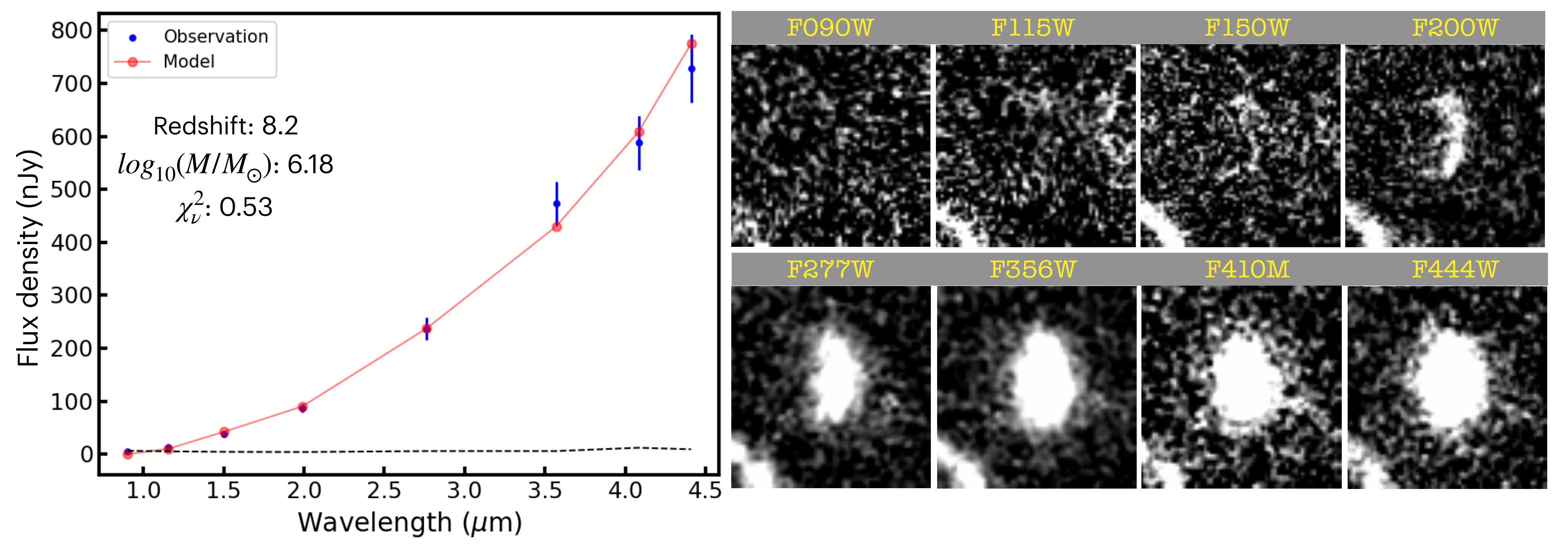}
\end{center}
\caption{DCBH candidates in PEARLS/NEP. In the left column, we show the JWST/NIRCam photometric fluxes in eight bands (blue dots) and the corresponding best-fitted photometric DCBH models (red dots and solid line) from \citet{Pacucci16}. The dashed lines correspond to NIRCam detection limits. In the right columns, we show the corresponding $2\times2$ arcsec images for the objects DCBH-1 (top) and DCBH-2 (bottom). Tabulated fluxes and coordinates of these objects are listed in Table~\ref{tab:fit-params}. }
\label{fig:candidates-SEDs}
\end{figure*}

The search for potential DCBH candidates among the 24,119 objects listed in PEARLS/NEP1$+$NEP2 catalog began by selecting observations with at least four detected fluxes above the JWST NIRCam 2$\sigma$ detection limit (see Sec.~\ref{sec:PEARLS}). For fluxes associated with unrealistically small error bars, such as $< 0.1$ mag, we capped both upper and lower error bars at 0.1 mag. An SED fitting using reduced-$\chi^{2}$ ($\chi^{2}_\nu$) minimization was then performed for every catalog object using all the \citet{Pacucci16} DCBH models described above. Consequently, two sources with IDs 21567 and 22802 revealed acceptable fits with $\chi^{2}_\nu < 3$ (hereafter, we refer to the two DCBH candidates as DCBH-1 and DCBH-2, designated in arbitrary order). Based on the minimum and maximum degrees of freedom (two and six), the corresponding limits of the P-values for the selected $\chi^{2}_\nu$ range are 0.049 and 0.006. Therefore, considering a maximum $\chi^{2}_\nu$ of three is quite generous for this analysis and ensures that the upper limits on the comoving density of DCBH host halos in Section~\ref{sec:limits} are conservative. The fit parameters and the source information of all detections are listed in Table~\ref{tab:fit-params}. In the case of DCBH-1, it appeared that the obtained mass ($\log_{10}(M/M_{\odot})$ = 5.56) falls below the range covered by our models. This happens because we chose to expand our model dataset by employing an extrapolation technique. We included additional models that were derived through interpolation, where we closely examined observed data points and created new models to fill in the gaps. By doing so, we significantly enhanced both the overall coverage, which means a broader range of scenarios, and the granularity within our model set. The mass of DCBH-1 was therefore scaled down in this process. Thumbnail images and SED fits of the two candidates are shown in  Figure~\ref{fig:candidates-SEDs}.

As expected, the photometric SEDs exhibit extremely red continuum slopes through the NIRCam bands. At the flux levels of these candidates, the first data point at wavelengths longward of the Ly$\alpha$ limit lies too close to the NEP detection threshold to allow a sharp Ly$\alpha$ limit feature to be seen (unlike in the $z=6$ and $z=10$  examples shown in Figure~\ref{fig:SED}). Instead, the continuum gradually fades below the detection limit, which unfortunately makes the redshift difficult to constrain. These photometric signatures may also be reproduced by obscured AGN or very dusty galaxies (see Sec.~\ref{fig:SED comparison} and Fig.~\ref{fig:degeneracy}). 

We note that there is a range of morphologies among these sources, with one of the candidates being extended and one being more point-like. Naively, one would expect DCBHs to appear as unresolved point sources, but the presence of ionized gas surrounding the DCBH and/or its proximity to a nearby galaxy could also be consistent with an extended source. However, one should note that if the extended source is indeed a galaxy, the \citet{Pacucci16} DCBH SED fitting method will not yield accurate results.

Shifting focus to a different aspect of the sources, it is noteworthy that the Eddington ratios in the models we used consistently demonstrate a slightly super-Eddington behavior \citep[as elucidated in Figure 1 of ][]{Pacucci15}. This characteristic is particularly evident in our two candidates, DCBH-1 and DCBH-2, which have estimated Eddington ratios of 3.1 and 1.5, respectively, at the time of observation. Notably, the estimate for the luminosity emitted by the black hole in GN-z11, the farthest detected to date \citep{Maiolino2023arXiv}, is about five times its Eddington luminosity.


\subsection{The comoving density of direct-collapse black holes} 
\label{sec:limits}

The prospects of detecting DCBHs in a given survey are set by a combination of two factors: (i) the flux detection limits of the photometric and/or spectroscopic observations used and (ii) the sky area covered by a survey. Whereas the flux detection thresholds set limits on the properties of the observable DCBHs (in terms of black hole mass, accretion rate, etc.), the survey area determines the number of such objects that are included in the survey. 

The connection between the survey area and the expected number of DCBH detections is further complicated by substantial uncertainties regarding the properties required for DCBHs to form. Theoretical predictions on the number density of halos that could host DCBHs at $z=6$--20 spans at least five orders of magnitude at any given redshift \citep{Habouzit16, Valiante17, Inayoshi_review_2019}. 

To compare current predictions on the number density of DCBH host halos to the number density that would allow for at least one detection in PEARLS, we derived the comoving number density $n_\mathrm{hosts}$ of host halos that would produce a certain number of observed DCBH candidates ($N_\mathrm{obs}$) in the $\Delta(z)$ interval centered around redshift $z$ as
\begin{equation}
n_\mathrm{hosts} = \frac{N_\mathrm{obs}}{V_\mathrm{c}(A,z,\Delta(z))} \frac{\Delta t_{z,\Delta(z)}}{\Delta t_\mathrm{DCBH}},
\label{eq:nhosts}
\end{equation}
where $V_\mathrm{c}(A,z,\Delta(z))$ is the comoving volume probed by a survey that covers an area $A$ in the sky across $\Delta(z)$ at redshift $z$, $\Delta t_{z,\Delta(z)}$ is the cosmic time interval spanned by $\Delta(z)$ at this redshift, and $\Delta t_\mathrm{DCBH}$ is the total time interval during which a DCBH can be expected to exhibit some specific set of selection criteria. 

In the case of the NEP, we based the $V_\mathrm{c}(A,z,\Delta(z))$ estimate on a survey area for two spokes (see below) and evaluated eq.~\ref{eq:nhosts} in $\Delta(z)=1$ bins. As long as $\Delta t_\mathrm{DCBH}<\Delta t_{z,\Delta(z)}$, $n_\mathrm{hosts}$ can be interpreted as the comoving number density of DCBH host halos emerging within a single redshift bin, which roughly corresponds to theoretical predictions on noncumulative densities of DCBH host halos \citep[e.g.][]{Habouzit_2016}.

By requiring DCBH candidates to be detectable in at least four JWST/NIRCam filters, the maximum redshift for which DCBHs may be detected throughout the full $0.5 - 7 \times 10^6\ M_\odot$ mass range considered corresponds to $z\approx 14$ when using PEARLS data for the NEP field. By comparison, other JWST cycle-1 surveys such as the Cosmic Evolution Early Release Science \citep[CEERS\footnote{\url{https://ceers.github.io}}; $\approx100$ arcmin$^{2}$ in the Extended Groth Strip;][]{Finkelstein2022Ceers} and the JWST Advanced Deep Extragalactic Survey \citep[JADES\footnote{\url{https://www.cosmos.esa.int/web/jwst-nirspec-gto/jades}}; $\approx175$ arcmin$^{2}$ in GOODS-S and GOODS-N;][]{Eisenstein23}, which have somewhat deeper detection limits, have the potential to extend this limit to $z\approx 16$ and $19$, respectively. 

\begin{figure}
\begin{center} 
\includegraphics[width=1.08\columnwidth]{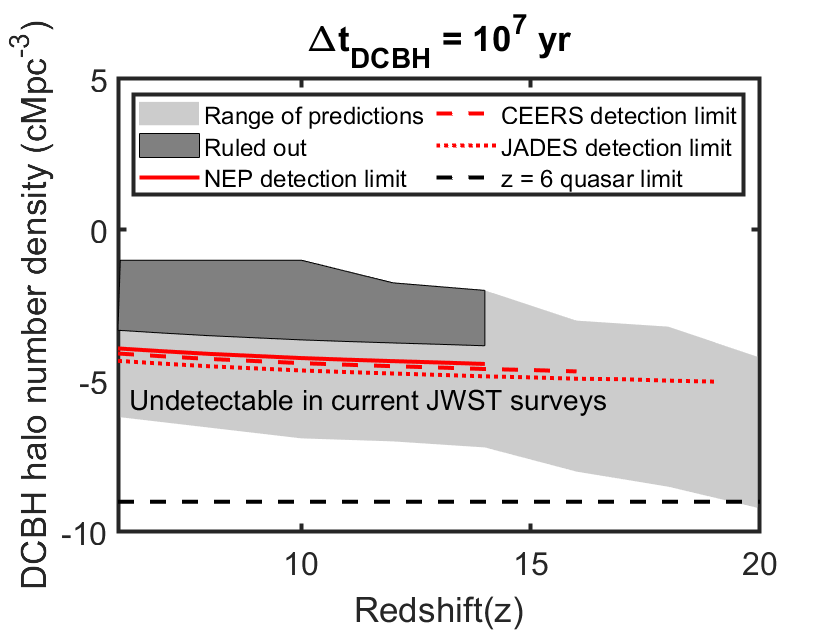}
\end{center}
\caption{Detection limits on the number density of halos that may host DCBHs as a function of redshift. The light-gray region represents the approximate range of theoretical predictions of the comoving number densities of halos that may host DCBHs, based on the compilation by \cite{Habouzit16}. The red lines represent the lowest detectable host halo comoving density in the case of a DCBH that remains sufficiently bright for detection in at least four JWST/NIRCam filters and retains its characteristic spectral signatures for $10$\,Myr given the total survey areas of PEARLS NEP (four spokes; solid red line), CEERS (dashed red) and JADES (dotted red). The part of the parameter space ruled out by the current limit derived in this paper is marked by the dark-gray region. The dashed black line represents the approximate $z=6$ quasar number density DCBH models must exceed prior to these redshifts to explain the SMBH powering these objects.}
\label{fig:comoving density} 
\end{figure}

By setting $N_\mathrm{obs}=1$ in equation~\ref{eq:nhosts}, Fig.~\ref{fig:comoving density} displays the minimum DCBH host halo number densities detectable in PEARLS NEP, CEERS, and JADES as a function of redshift under the assumption that $0.5 - 7 \times 10^6\ M_\odot$ DCBHs retain their \citet{Pacucci16} telltale spectral signatures for $\Delta t_\mathrm{DCBH} = 10$\,Myr. 

These detection limits were compared to the range of theoretical predictions on the comoving number densities of DCBH hosts from \cite{Habouzit16}. The DCBH models that fall above these limits could produce DCBH detections within the specified survey areas, whereas models falling below the limits would produce too few DCBHs per comoving volume to make detections likely.

Based on our analysis using PEARLS NEP1 and NEP2 observations, we found two potential sources at different redshifts (see~Sec.~\ref{sec:candidate}~for details). Assuming $N_\mathrm{obs}=2$ and the PEARLS survey area of $\sim34$\,arcmin$^2$ for two spokes, the upper limit on the comoving density of halos that host DCBHs that have grown to $5\times 10^5$--$7\times 10^6\ M_\odot$ and exhibit SEDs in accord with the \citet{Pacucci16} models then becomes $\approx 1.5$--5$\times 10^{-4}$\,cMpc$^{-3}$ ("c" stands for comoving density; the lower part of the dark-gray region in Fig.~\ref{fig:comoving density}) at $z=6-$14. These limits are conservative since they were derived under the assumption that both candidates could potentially be located at any redshift in this range. We stress that the redshift estimates for the candidates obtained by \citet{Pacucci16} are photometric and therefore subject to substantial uncertainties (as also discussed in Sec.~\ref{sec:SED}).

These limits already exclude the upper part of the DCBH parameter space compiled by \citet{Habouzit16} at these redshifts, and increasing the timescale over which the DCBHs are assumed to be detectable given our search criteria would strengthen the constraints even further. For example, increasing $\Delta t_\mathrm{DCBH}$ from 10 Myr to 100 Myr would shift both the upper limit and the theoretical detection limits of the various surveys down by 1 dex at $z\approx 6$--8. At higher redshifts, the limits could potentially become even stronger, but scenarios of this type (with $\Delta t_\mathrm{DCBH}>\Delta t_{z,\Delta(z)}$ for $\Delta(z)=1$, causing DCBHs that form in higher-redshift bins to survive into lower-redshift bins) would result in upper limits that depend on the formation history of DCBH host halos specific to each formation model.

Since the predicted range of host number densities extends well above the detection limits of PEARLS/NEP, CEERS, and JADES, DCBH detections may be possible even if a small fraction of potential DCBH host halos would produce sources with the spectral characteristics we used for candidate selection. However, since the predicted range also extends below the detection limits, success in detecting DCBHs in these JWST surveys is not assured. In Fig.~\ref{fig:comoving density}, we considered only number densities from the \cite{Habouzit16} simulation scenarios that successfully generate DCBH candidate halos, not the \cite{Habouzit16} simulation scenarios that failed to produce any such objects within the simulated volume. If the latter actually provides a better representation of reality, then the DCBH number density could in principle be even lower than what is shown in Fig.~\ref{fig:comoving density}. However, if DCBHs serve as the primary seeds of early SMBHs, a hard lower limit can be set by considering the comoving number densities of $z\approx 6$ quasars ($\sim 10^{-9}$\,cMpc$^{-3}$; e.g., \citealt{Valiante17}) since the DCBH host halo number density must then exceed this limit at some prior redshift. 

\begin{table*}[htbp]
\caption{List of the selected sources and the corresponding best-fit parameters obtained with the $z>6$ DCBH models \citep{Pacucci16}.}
\label{tab:fit-params}
\centering
\vspace*{\fill}
\begin{tabular}{lcc}
\hline
\hline

Parameters/ObsID$^{a}$  & DCBH-1       & DCBH-2                        \\
\hline
RA(deg)             &  260.75396245   & 260.69140947          \\
Dec(deg)            &  $+$65.80118827 & $+$65.79172734        \\ 

F090W               & --              & 29.59$\pm$1.62      \\
F115W               & --              & 28.57$\pm$0.62                  \\
F150W               & --              & 27.43$\pm$0.19    \\
F200W               & 29.91$\pm$0.79  & 26.55$\pm$0.10$^{b}$      \\ 
F277W               & 28.83$\pm$0.15  & 25.46$\pm$0.10    \\ 
F356W               & 28.42$\pm$0.10  & 24.71$\pm$0.10   \\ 
F410M               & 27.95$\pm$0.13  & 24.48$\pm$0.10   \\ 
F444W               & 27.50$\pm$0.10  & 24.24$\pm$0.10    \\ 

Redshift            & 14.1            & 8.2               \\ 
$\log_{10}(M/M_{\odot})$              & 5.56            & 6.18   \\
Eddington ratio$^{c}$     & 3.1             & 1.5                  \\
$\chi^{2}_{\nu}$    & 2.41            & 0.53         \\
\hline
\toprule

	 & 	 
\end{tabular}

\tablefoot{
$^{a}$Observation ID based on the combined NEP1 and NEP2 catalog. $^{b}$Error bars smaller than 0.1 mag were set to 0.1 mag. $^{c}$ Eddington ratio obtained from the SED fitting for the two DCBH candidates (see the text).
}

\end{table*}


\subsection{Constraints on X-ray emission}
\label{sec:X-ray}
In order to constrain the X-ray emission of our two candidates (see Fig. \ref{fig:candidates-SEDs}), we used the 1.3-megasecond \ch\ observations available in the field. By using different extraction regions centered at each source position (the 50\% encircled energy, 0.32 arcsec in all cases, and $r = 1$\,arcsec) and 3"--5" annulus for background extraction, we obtained nondetections for both of the candidates. In this case, we could compute 3$\sigma$ upper limits on count rates of $1.7\times 10^{-6}$\,count\,s$^{-1}$, which translates into an upper limit on the observed flux of $\sim 3 \times 10^{-17}$ \flux\ in the 0.5--7\,keV band when assuming a power-law model with a spectral index of $\Gamma=1.4$. This flux translates into an X-ray luminosity of $\sim 7 \times 10^{43}$ in the 0.5--7\,keV band at redshift $z=7$. The {\it XMM-Newton} and {\it NuSTAR} data available in this field \citep{zhao21} were also searched at the position of the two sources, and a consistent upper limit with \ch\ was found. 
We note that the two DCBH candidates selected by \cite{Pacucci16} are strong detections, with a full-band X-ray flux of $\sim 6-7 \times 10^{-16}$ \flux calculated from $125$\,counts.

\section{Discussion and conclusions}
\label{sec:discussion}
In this study, we conducted an observational search for DCBHs, which represent one of the proposed mechanisms for the formation of SMBHs. The DCBH scenario elucidates the process by which black holes achieved masses exceeding 10$^{9} \Msun$ during the early stages of the Universe. With the launch of JWST, we now possess the capability to explore the depths of high redshifts and low luminosities, enabling us to delve into the past and investigate this phenomenon with unprecedented precision in the IR regime. 

As we have shown, objects with photometric SEDs that closely resemble the \cite{Pacucci16} predictions of $z\gtrsim 6$ DCBHs do exist, even in full eight-band JWST/NIRCam datasets of the type provided by the PEARLS/NEP survey. Consequently, two DCBH candidates, a point-like source, and an extended source were identified through SED fitting using the \cite{Pacucci16} DCBH models (see~Sec.~\ref{sec:candidate} for details).

However, since a DCBH sample selected in this way may include heavily dust-reddened galaxies and AGN (see Sect.~\ref{sec:SED}), additional data were required to ascertain the true nature of these objects. Given the ubiquity of dust-obscured AGN at $z>5$ revealed by JWST \citep{Kocevski2023,Greene23}, it is possible that DCBH candidates identified by our search criteria are in reality AGN for which the flatter part of the SED is simply hiding below the detection threshold in the shorter-wavelength filters. Follow-up spectroscopy is therefore required to elucidate the true nature of objects that display DCBH-like photometric SEDs.

The spectroscopic signatures of the DCBH models by \citet{Pacucci16} include very strong hydrogen Balmer lines, an H$\alpha$-H$\beta$ emission-line ratio of H$\alpha$/H$\beta>10$ (i.e., in significant excess of the case B recombination value of $\approx 2.8$ due to collisional pumping of H$\alpha$), a strong HeII1640 emission line, and an absence of emission lines due to metals \citep{Pacucci_2017_CR7}. At $z\gtrsim 6.9$, H$\alpha$ redshifts out of the JWST/NIRSpec wavelength range, but it remains within reach of JWST/MIRI imaging and spectroscopy until $z\approx 15$. Follow-up observations with NIRSpec and MIRI could therefore rule out both dusty starbursts and AGN, at least for the brightest of our candidates.

If the DCBH is located inside a large ionized bubble in the intergalactic medium (for instance because the formation of the DCBH has been triggered by a nearby AGN; e.g., \citealt{Johnson19}), the Ly$\alpha$ line could also be partially transmitted. However, the transmission of a significant Ly$\alpha$ flux at $z\gtrsim 6$ would alter the photometric signature \citep[e.g.,][]{Zackrisson11b}. The DCBH candidate selection in this work and by \citet{Pacucci16} were made assuming a negligible Ly$\alpha$ contribution to the SED.

Due to stochastic gas accretion onto the black hole, DCBHs may also display brightness variations \citep{Wang_2017} and thus stand out in multi-epoch observations with NIRCam. While this makes DCBHs distinct from dusty galaxies (except in the rare case of a supernova going off in a star-forming galaxy), it does not, however, effectively separate DCBHs from AGNs. 

The extended morphology evident in one of our candidates could arise from ionized gas around the DCBH, as shown in some early simulations. However, the morphological structure at least shortly after their formation process would be nearly spherical in this case, with a typical radius of $0.5$ kpc \citep[see, e.g.,][]{Latif2013}. Specifically, DCBH-2 shows a somewhat extended morphology, with a maximum size of a few kiloparsecs. The size of the luminous, possibly ionized region depends on the frequency range observed by the filter, ranging from a fraction of a kiloparsec to a few kiloparsecs. Detailed simulations are needed to assess how the optical depth of this ionized region depends on the observed wavelength.

If the extended structure is not associated with the central DCBH, it must be caused by either the DCBH host galaxy or a nearby galaxy. However, an evident stellar component in the candidate's image would substantially impact the SED, invalidating the use of DCBH models without accounting for the galaxy contribution. Consequently, our current search approach is designed to avoid detecting such objects. As this paper focuses on identifying DCBHs in their "pure'' (or nearly pure) state, the upper limits on the comoving density of DCBH host halos that we present can admittedly be circumvented in scenarios where star formation is quickly triggered in the host halos of newborn DCBHs, where the DCBH host halo merges with a nearby star-forming halo or in which the DCBH SED otherwise deviates significantly from the \citet{Pacucci16} prediction.


\begin{acknowledgements}
AN and EZ acknowledge funding from Olle Engkvists Stiftelse. EZ also acknowledges grant 2022-03804 from the Swedish Research Council, funding from the Swedish National Space Agency and has benefited from a sabbatical at the Swedish Collegium for Advanced Study. WPM acknowledges that the National Aeronautics and Space Administration provided support for this work through Chandra Award Numbers GO8-19119X, GO9-20123X, GO0-21126X, and GO1-22134X issued by the Chandra X-ray Center, which the Smithsonian Astrophysical Observatory operates for and on behalf of the National Aeronautics Space Administration under contract NAS8-03060. FP acknowledges support from a Clay Fellowship administered by the Smithsonian Astrophysical Observatory. This work was also supported by the Black Hole Initiative at Harvard University, which is funded by grants from the John Templeton Foundation and the Gordon and Betty Moore Foundation. 
This work is also based on observations made with the NASA/ESA/CSA James Webb Space
Telescope. The data were obtained from the Mikulski Archive for Space
Telescopes at the Space Telescope Science Institute, which is operated by the
Association of Universities for Research in Astronomy, Inc., under NASA
contract NAS 5-03127 for JWST. These observations are associated with JWST
programs 1176 and 2738.
RAW, SHC, and RAJ acknowledge support from NASA JWST Interdisciplinary
Scientist grants NAG5-12460, NNX14AN10G, and 80NSSC18K0200 from GSFC. Work by
CJC and NJA acknowledge support from the European Research Council (ERC) Advanced
Investigator Grant EPOCHS (788113). BLF thanks the Berkeley Center for
Theoretical Physics for their hospitality during the writing of this paper.
MAM acknowledges the support of a National Research Council of Canada Plaskett
Fellowship, and the Australian Research Council Centre of Excellence for All
Sky Astrophysics in 3 Dimensions (ASTRO 3D), through project number CE17010001.
CNAW acknowledges funding from the JWST/NIRCam contract NASS-0215 to the University of Arizona.
AZ acknowledges support by Grant No. 2020750 from the United States-Israel Binational Science Foundation (BSF) and Grant No. 2109066 from the United States National Science Foundation (NSF), and by the Ministry of Science \& Technology, Israel.
We also acknowledge the indigenous peoples of Arizona, including the Akimel
O'odham (Pima) and Pee Posh (Maricopa) Indian Communities, whose care and
keeping of the land has enabled us to be at ASU's Tempe campus in the Salt
River Valley, where much of our work was conducted.

\end{acknowledgements}

\bibliographystyle{aa}
\bibliography{library}

\begin{appendix}
\section{Properties of galaxies with the potential to mimic the spectral energy distributions of direct-collapse black holes}
\label{Appendix-A}
In this appendix, we provide additional constraints on the properties of galaxies capable of mimicking the photometric signatures of DCBHs. Treating the \citet{Pacucci16} DCBH SED models as mock observations and attaching realistic observational errorbars to the resulting NIRCam photometric data points, we used the \textsc{Bagpipes} code \citep{Carnall18} to fit galaxy models to these data. We found that the mock NIRCam photometry can be reproduced by galaxies with a dust attenuation of $A_V \gtrsim 2$ mag over a wide range of redshifts.

\begin{figure*}
\begin{center}
\includegraphics[width=1.7\columnwidth]{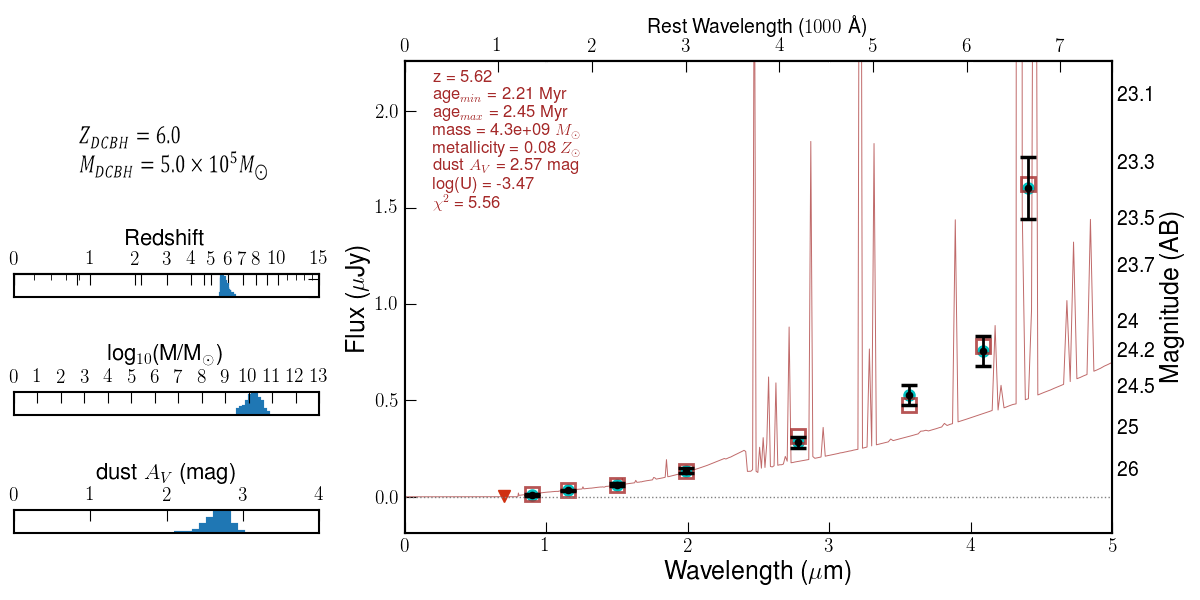}
\includegraphics[width=1.7\columnwidth]{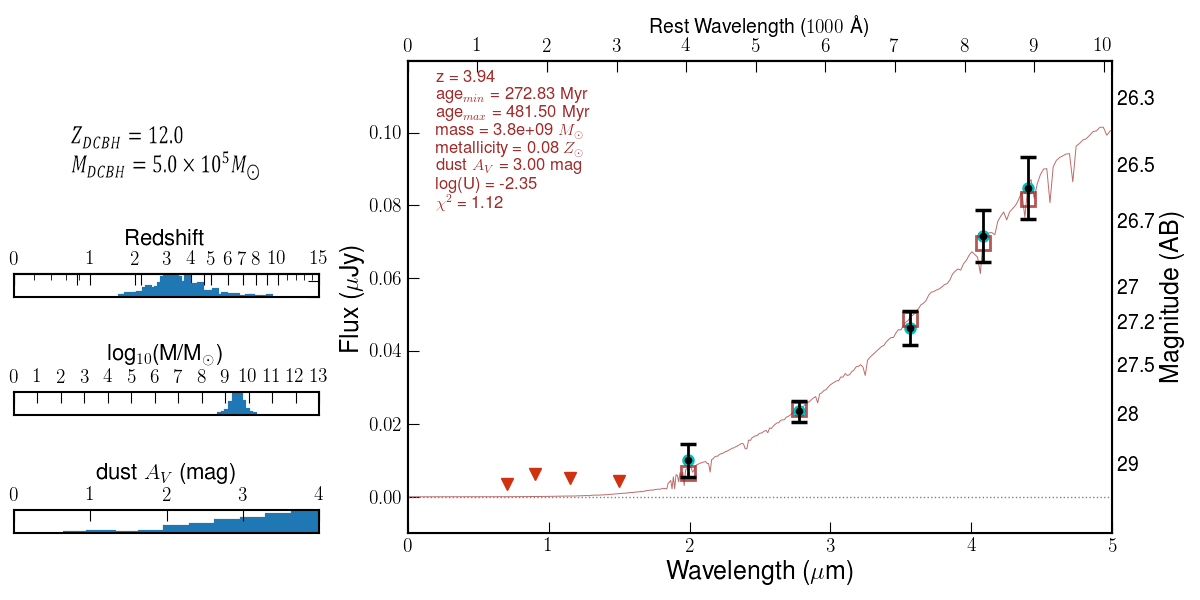}
\end{center}
\caption{\textsc{Bagpipes} fits of galaxy models to mock DCBH data. The red squares represent the best-fitting galaxy model, based on the red model galaxy spectrum, to the photometric fluxes of  $M_\mathrm{DCBH}=5\times 10^5\ M_\odot$ DCBH model (black circles with mock error bars) from  
\citet{Pacucci16}. Red triangles indicate filters where the mock DCBH fluxes fall below the PEARLS detection limits of the NEP field. The estimated errors on the fitted redshift, total stellar mass, and dust attenuation of the galactic fit are indicated by the sliders to the left of each plot. \textit{Top:} Example where a $z_\mathrm{DCBH}=6$ spectrum is fitted by a dusty galaxy model at a similar redshift ($z\approx 5.62$). \textit{Bottom:} Example where a $z_\mathrm{DCBH}=12$ spectrum is fitted by a dusty galaxy model at a significantly lower redshift ($z\approx 3.94$).   
\label{fig:degeneracy}}
\end{figure*}

Figure~\ref{fig:degeneracy} presents two cases where mock NIRCam data generated from DCBH models at $z_\mathrm{DCBH}=6$ (a) and $z_\mathrm{DCBH}=12$ (b) DCBH are fitted by models involving highly dust-reddened stellar populations. The first case involves a dusty stellar population at a redshift similar to that of the DCBH model, while the second case involves a dusty galaxy at a significantly lower redshift of $z\approx 4$. It is also possible to find solutions in which the best-fitting galaxy models have redshifts higher than those of the underlying DCBH models. However, since this requires extremely dusty galaxies that form very early in the history of the Universe, such interlopers should be considerably less common \citep{Algera_2023}. Moreover, medium-band filters are also effective at removing dusty galaxies.

\end{appendix}

\end{document}